\documentclass[a4paper]{jpconf}
\usepackage[utf8]{inputenc}
\usepackage{graphicx}
\usepackage{iopams}
\begin{document}
\title{Liquid tin droplet fragmentation by ultra-short laser pulse}

\author{S~Yu~Grigoryev,$^{1,2}$ S~A~Dyachkov,$^{1,2,3}$ V~A~Khokhlov,$^2$ V~V~Zhakhovsky,$^{1,2}$ A~N~Parshikov,$^1$ N~A~Inogamov$^{2,1}$}
\address{Dukhov Research Institute of Automatics (VNIIA), ul. Sushchevskaya 22, Moscow 127055, Russia}
\address{Landau Institute for Theoretical Physics of the Russian Academy of Sciences, Akademika Semenova 1a, Chernogolovka, Moscow Region 142432, Russia}
\address{Joint Institute for High Temperatures of the Russian Academy of Sciences, ul. Izhorskaya 13 Bldg~2, Moscow 125412, Russia}

\ead{nailinogamov@gmail.com}

\begin{abstract}
The fragmentation of a liquid metal droplet induced by a nanosecond laser pulse has been studied well. However, the fragmentation mechanism may be different, when a sub-picosecond laser pulse is applied. To discover the details of the fragmentation process, we perform a hydrodynamic simulation of a liquid tin droplet irradiated by a femtosecond laser pulse. We have found that the pressure pulse induced by an instantaneous temperature growth in the heated layer propagates from the one side of the surface of a spherical droplet and focuses in its center; at the release a big cavity is formed at the center of a droplet; the pressure wave release at the backside surface may cause the spallation.
\end{abstract}

\section{Introduction}

The laser induced plasma from the surfaces of irradiated tin targets is the promising light source for the extreme ultraviolet lithography \cite{wagner:2010, sullivan:2015,medvedev:2018}. However, the formation of plasma is accompanied by the debris consisting of the secondary droplets of tin. This cloud may enhance the absorption of light, so that the discovery of fragmentation mechanisms, leading to the specific spatial and size distributions of fragments, is of practical interest.

The droplet fragmentation process is known to be dependent upon laser pulse duration. Thus, a nanosecond pulse leads to a surface ablation with production of plasma, which applies a recoil pressure to a droplet and deforms it until eventual fragmentation\cite{klein:2015, gelderblom:2016,kurilovich:2016}. A picosecond or femtosecond laser pulse changes the process qualitatively: such short laser impact turns a droplet into a hollow structure surrounded by a liquid metal shell which fragments during the expansion process~\cite{vinokhodov:2016,krivokorytov:2017,grigoryev2018expansion}. Understanding of fragmentation mechanism induced by a sub-picosecond laser pulse requires a complex physical model. To shed a light on this problem we perform the complete three-dimensional (3D) hydrodynamic simulation of a liquid tin droplet irradiated by a femtosecond laser pulse. The smoothed particle hydrodynamics (SPH) method \cite{1985:CPR:Monaghan} seems to be the most suitable for this kind of phenomena: it naturally handles the complex boundaries, which appear during the cavitation and spallation. The Riemann problem solution is applied to an interparticle contact \cite{2002:JCP:Parshikov} for the proper pressure and velocity calculation what results in stable solutions for shock and release waves. We also use the efficient software tools for large-scale parallel simulations \cite{2017:LJM:Dyachkov}.

\section{Simulation approach and setup}

\begin{figure*}[t]
\center{
\includegraphics[width=1.0\linewidth]{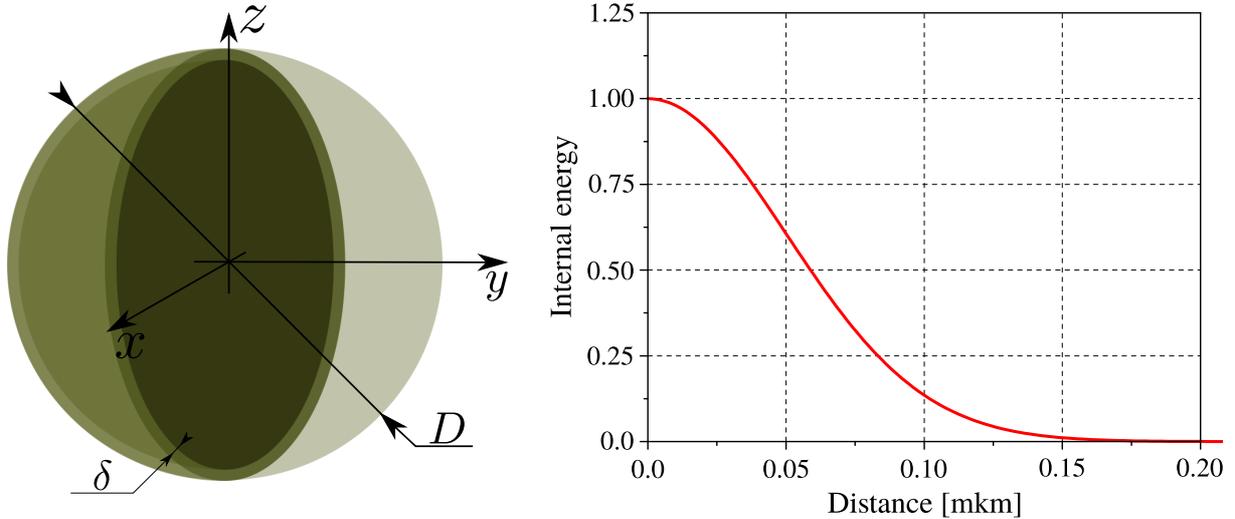}
\caption {\label{fig:droplet} Left: liquid tin, the drop diameter is $D = 2~\mu$m, the thickness of the heated layer is $\delta \sim 100$~nm, $y$ is the direction of the laser beam. The energy of laser radiation is absorbed by the left hemisphere surface (bright color). Right: internal energy distribution at the initial moment.}
}
\end{figure*}

The representation of a continuous medium with smoothed particles has many advantages. Its meshless nature does not require special algorithms to handle contact surfaces and internal or external free boundaries. Thus it allows reproducing of the cavitation and spallation processes naturally. 

A simulated sample is represented by a set of SPH particles having the material properties. In the experiments the sizes of liquid metal droplets are usually several tens of microns. However, the heated layer $\delta$ is quite thin (about $100$~nm) what defines the required spatial resolution for a 3D SPH droplet representation. To resolve this layer properly we need at least 10 particles what results in the particle size of $10$~nm. For example, to represent a droplet of the radius $R = 50$~$\mu$m we need a half of a billion particles what is rather expensive and difficult to simulate. Instead, we use the droplet of the radius $R=1$~$\mu$m (figure \ref{fig:droplet}) assuming the main fragmentation features to be unchanged.

The numerical simulation of a liquid tin drop as continuous medium utilizes the following conservation laws for the mass, the momentum, and the energy:
\begin{equation}
\label{eq:continuity}
\dot{\rho} + \nabla \cdot \mathbf{U}=0,
\end{equation}
\begin{equation}
\label{eq:momentum}
\rho \dot{\textbf{U}} + \nabla P=0,
\end{equation}
\begin{equation}
\label{eq:energy}
\rho \dot{E} + \nabla \cdot \left(P \mathbf{U}\right) = 0.
\end{equation}
The total specific energy $E$ in the equation (\ref{eq:energy}) consists of the inner and kinetic parts ($E=e+\mathbf{U}^2/2$); $\rho$~and $\mathbf{U}$ are the density and the velocity respectively, $P$ is the hydrostatic pressure. The system of equations (\ref{eq:continuity})-(\ref{eq:energy}) is enclosed by the equation of state which provides a connection between the pressure $P$, the density $\rho$, and the inner specific energy $e$:
\begin{equation}
\label{eq:Grun}
P= P(\rho,e)= P_r + \Gamma\rho(e - e_r),
\end{equation}
where $\Gamma$ is the Gruneisen parameter, where $P_{r}$, $e_{r}$ are the reference pressure and the specific energy for the shock Hugoniot in the form $u_{s}=c+au_{p}$, $u_{s}$ is the shock velocity, $u_{p}$ is the particle velocity, $c$ is the bulk sound speed, $a$ is the coefficient:
\begin{equation}
\label{eq:eos_reference}
P_r(x) = \rho_0 c^2 \frac{1 - x}{[1 - a(1 - x)]^2}, \qquad e_r(x) = \frac{P_r}{\rho_0}\frac{1 - x}{2}, \qquad x = \rho_0/\rho.
\end{equation}

The system of equations (\ref{eq:continuity})-(\ref{eq:Grun}) is solved using the SPH method. The particles interact with each other within the smoothing distance controlled by a kernel function. At the interparticle contacts the Riemann solutions for the pressure and the velocity are applied. The hydrostatic pressure for liquid tin is calculated according to the Mie-Gruneisen equation of state. The detailed description of contact SPH approach is introduced in \cite{2002:JCP:Parshikov}. The summary of equation of state parameters for the SPH-simulated liquid tin are given in the table~\ref{table:tin}.

\begin{table}
\center{
\caption{\label{table:tin} Liquid tin properties. }
\label{tab:a}
\begin{tabular}{ll}
\hline
    {Property}                             & {Value} \\
\hline
\hline
Bulk modulus $B,$ GPa                        & 48.9 \\
Normal density $\rho_{0},$ kg/m$^{3}$        & 7287  \\
Specific heat capacity $C_v$, J/(kg$\cdot$K)  & 255   \\
Tensile strength $T$, GPa                 & 1.0   \\
Gruneisen parameter $\Gamma$                 & 1.7   \\
Shock hugoniot coefficient $c,$ km/s       & 2.59  \\
Shock hugoniot coefficient $a$             & 1.49  \\
\hline
\end{tabular}
}
\end{table}

\section{Surface heating}

The well known fact is that a femtosecond laser pulse is absorbed by free electrons on a metal surface during $10$~fs, i.e. almost instantaneous as compared to the pulse duration of $800$~fs. The energy is gradually transferred to the ionic lattice owing to the electrons relaxation. Within the $\sim 10$~ps period, the temperatures of electrons and ions reach equilibrium. 

Our simulation does not account for procesess taking place in the first $10$~ps of irradiation: we model a laser pulse absorption by setting a specific inner energy in a skin layer of a droplet using the following expression:
\begin{equation}
\label{eq:energI-3D}
e(r, \theta) = e_{0} \exp\left[-\frac12\left(\frac{r-R}{\alpha \delta}\right)^2\right] H(\pi / 2 - \theta)\cos\theta
\end{equation}
Here $e_{0}$ is the peak amplitude, $R$ is the droplet radius, $\delta$ is the skin layer thickness, $\alpha$ is the absorption decay coefficient, $r$ is the radial position of a point within a droplet, $\theta$ is the angle between the laser beam and a point within a droplet. The peak amplitude $e_0$ was varied from 1 to 5~MJ/kg, because it covers the two observed in the experiments  different droplet fragmentation regimes. The $\alpha$ coefficient has no direct physical interpretation, but it was chosen from the assumption that the liquid tin absorbs radiation energy within a near-surface layer of thickness $\delta \sim 100$~nm. We used for the $\alpha$ value 1.44$\times$10$^2$. With this value, the internal energy distribution shown in figure 1 is obtained. The parameters are adjusted so that the peak pressure amplitude is about 10~GPa. 

The Gauss distribution of the absorbed energy near the surface of a droplet seems to be a good approximation. The electrons on the surface have greater kinetic energy what results in a more intense heating of ions. However, the heating gradually decreases with the depth due to less energy absorbed by electrons.

The dependency on the angle $\theta$ is introduced to account the anisotropic heating of electrons within the skin layer. As soon as the boundary conditions for incident and reflected beams change with an angle, the absorption is maximal for $\theta = 0$ and minimal for $\theta = \pi/2$, what is properly reproduced by the $\cos\theta$ function. Finally, the Heaviside function $H(\pi / 2 - \theta)$ defines the laser irradiation for a half of the droplet.

\section{Spallation/cavitation model}

A sub-picosecond laser irradiation heats a thin surface layer of a liquid tin droplet what results in an instantaneous temperature growth and pressure increase. Next, the short pressure pulse propagates from a heated surface to a center what is demonstrated in the next section. The pulse is followed by a rarefaction wave which causes the material extension. As the stress in material reaches the critical value of the tensile strength it relaxes to zero with the formation of free surfaces with the cavities and spalls in between. The tensile strength not only depends on a material state defined by the temperature and the pressure, but also on dynamic properties such as the strain rate.

Most solid materials are known to have the tensile strength that decreases with the growth of the temperature and drops dramatically in a liquid state. Thus, at the low strain rate $5.0\times10^{4}$~s$^{-1}$ the tensile strength of solid tin is about 1.2~GPa. In a liquid state it decreases by an order of magnitude from 1.2 to 0.12~GPa\cite{kanel:2015}. 

The strain rate is known to affect the tensile strength as well \cite{ashitkov:2016}: at relatively high strain rates about $1.3\times10^{9}$~s$^{-1}$ the tensile strength of liquid tin increases up to 1.9~GPa. The theoretical tensile strength limit is known to be about 6~GPa which is hardly achieved in an experiment. In our simulations the strain rate is of the order $10^{8}$~s$^{-1}$, so we use the intermediate value $1.0$~GPa for the tensile stength.

Simulation of spallation or cavitation processes is conducted using the following simple model. Each pair of SPH particles providing the contact Riemann solution for stress, which exceeds the tensile strength, lose their cohesion. That condition results in the almost instantaneous relaxation of the tensile stress and formation of a cavity between the particles.

\section{Simulation results}

\begin{figure*}[t]
\center{
\includegraphics[width=1.0\linewidth]{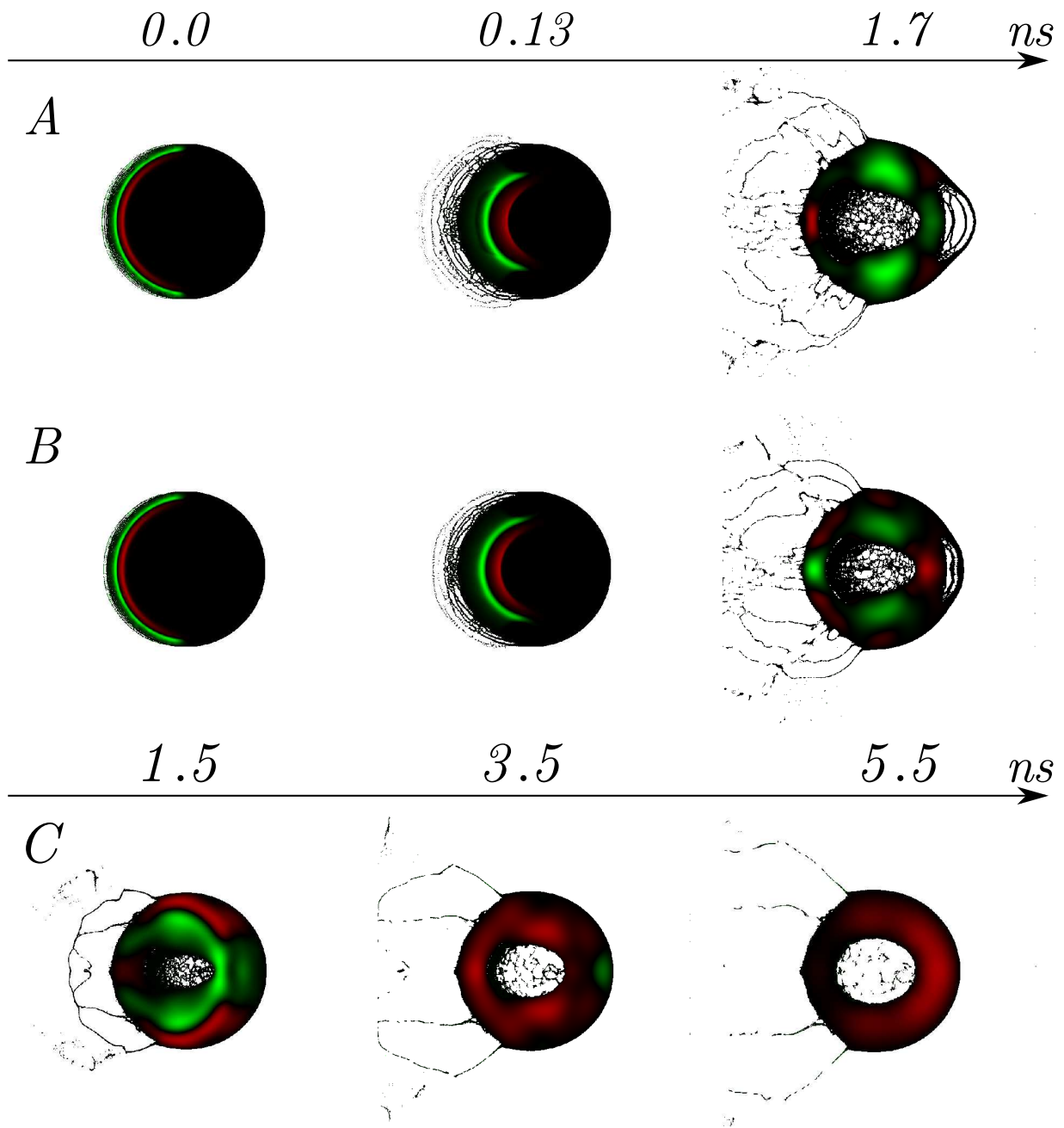}
\caption{\label{fig:3D-maps} The droplet fragmentation evolution. $(a)$ the high laser intensity, the considerable cavity bubble at the center and the backside spall formation; $(b)$ the medium laser intensity, the smaller bubble and spall; $(c)$ the small laser instensity, the complete inhibition of the spall, the bubble still exists.}
}
\end{figure*}

The evolution of the liquid tin droplet after laser irradiations with various intensities is presented in figure~\ref{eq:energI-3D} using two-dimensional color maps for the pressure. The maps are built within the plane which includes the laser beam propagating from the left to the right and the center of the droplet. The red color corresponds the compression ($P>0$) while the green color is for the rarefaction ($P < 0$). The laser intensity changes from high $(a)$ to low $(c)$.

The high laser intensity regime, shown in the figure \ref{eq:energI-3D}$(a)$, is accompanied by the instantaneous pressure growth to the magnitude of several 10~GPa in the heated layer of the droplet. The pressure pulse unloads to a free surface what results in the heated layer extension and ablation. It should be noted that we do not include liquid-gas or liquid-plasma phase transition to the equation of state of tin. The velocity of the ablated material is maximal along the laser beam direction (equator, $\theta = 0$) and reduces to the poles ($\theta = \pi/2$) what corresponds the initial energy deposition \eqref{eq:energI-3D}.

The initially narrow pressure pulse begins to propagate from the heated layer to the center of the droplet. After $130$~ps the compression wave is widened (red half-moon), it is followed by the rarefaction wave (green half-moon), and the peak pressure is reduced. However, at some moment the peak pressure begins to increase near the center of the droplet due to energy cumulation. Similarly, the following rarefaction wave focuses at the center of the droplet what results in the extreme extension which exceeds the tensile strength. At this moment our cavitation model relaxes the tensile stresses what results in quite large cavity bubble growth in the center of the droplet.

As soon as the pressure pulse passes the center of the droplet its amplitude again begins to decrease. It propagates to the backside surface and releases. In this case of the high intense laser irradiation the amplitude of the pressure pulse is enough to form a rarefaction wave with tensile stresses which exceeds the tensile strength and forms spalls near the backside surface of the droplet. The spallation zone width is about $\sim 200$~nm, and the most fragmentation is located along the laser beam. 

The results with the two times smaller laser intensity $e_{0}$ from \eqref{eq:energI-3D} are shown in the figure~\ref{fig:3D-maps}$(b)$. Again, the skin layer heating results in its ablation; however the velocity of fragments is reduced. The cavity bubble in the center of the droplet exists, but its size is shortened. The spallation at the backside surface also takes place, but it is less intense than in the figure~\ref{fig:3D-maps}$(a)$. The spallation is completely inhibited in the simulation with the again reduced energy deposition shown in the figure \ref{fig:3D-maps}$(c)$. The intensity of the pulse seems to be not enough to produce a rarefaction wave from the backside surface exceeding the tensile strength. But the bubble in the center of the droplet still exists: the extension is considerably amplified by the spherical convergence of waves.

\section{Conclusion}

The hydrodynamic simulation of the liquid tin droplet fragmentation induced by an ultrashort laser pulse demonstrates two fragmentation mechanisms. First, the cavity bubble is formed in the center of a droplet due to rarefaction wave focusing. Second, the spallation near the backside surface of a droplet may occur at high enough pulse intensities.

\ack
The authors gratefully thank Slava Medvedev for stating the problem and valuable discussions.

\section*{References}
\bibliographystyle{iopart-num}

\begin{thebibliography}{10}
\expandafter\ifx\csname url\endcsname\relax
  \def\url#1{{\tt #1}}\fi
\expandafter\ifx\csname urlprefix\endcsname\relax\def\urlprefix{URL }\fi
\providecommand{\eprint}[2][]{\url{#2}}

\bibitem{wagner:2010}
Wagner C and Harner N 2010 {\em Nature Photonics\/} {\bf 4} 24--26

\bibitem{sullivan:2015}
O'Sullivan G, Li B, D'Arcy R, Dunne P, Hayden P, Kilbane D, McCormack T, Ohashi
  H, O'Reilly F, Sheridan P {\em et~al\/} 2015 {\em J. Phys. B\/} {\bf 48}
  144025

\bibitem{medvedev:2018}
Medvedev V~V, Grushin A~S, Krivtsun V~M, Vinokhodov A~Y, Antsiferov P~S,
  Krivokorytov M~S, Astakhov D~I, Abramenko D~B, Dorokhin L~A, Snegirev E~P,
  Yakushev O~F, Lakatosh B~V, Gayazov R~R, Solomyannaya A~D, Tsygvintsev I~P,
  Kim D~A, Sidel'nikov Y~V, Yakushkin A~A, Lash A~A, Ryabtsev A~N and Koshelev
  K~N 2018 {\em Phys. Usp.\/}  61

\bibitem{klein:2015}
Klein A, Bouwhuis W, Visser C~W, Lhuissier H, Sun C, Snoeijer J~H, Villermaux
  E, Lohse D and Gelderblom H 2015 {\em Phys. Rev. Appl.\/} {\bf 3} 044018

\bibitem{gelderblom:2016}
Gelderblom H, Lhuissier H, Klein A~L, Bouwhuis W, Lohse D, Villermaux E and
  Snoeijer J~H 2016 {\em J. Fluid Mech.\/} {\bf 794} 676--699

\bibitem{kurilovich:2016}
Kurilovich D, Klein A~L, Torretti F, Lassise A, Hoekstra R, Ubachs W,
  Gelderblom H and Versolato O~O 2016 {\em Phys. Rev. Appl.\/} {\bf 6} 014018

\bibitem{vinokhodov:2016}
Vinokhodov A~Y, Koshelev K~N, Krivtsun V~M, Krivokorytov M~S, Sidelnikov Y~V,
  Medvedev V~V, Kompanets V~O, Melnikov A~A and Chekalin S~V 2016 {\em Quantum
  Elec.\/} {\bf 46} 23

\bibitem{krivokorytov:2017}
Krivokorytov M~S, Vinokhodov A~Y, Sidelnikov Y~V, Krivtsun V~M, Kompanets V~O,
  Lash A~A, Koshelev K~N and Medvedev V~V 2017 {\em Phys. Rev. E\/} {\bf 95}
  031101

\bibitem{grigoryev2018expansion}
Grigoryev S~Y, Lakatosh B, Krivokorytov M, Zhakhovsky V, Dyachkov S, Ilnitsky
  D, Migdal K, Inogamov N, Vinokhodov A~Y, Kompanets V {\em et~al\/} 2018 {\em
  Phys. Rev. Appl.\/} {\bf 10} 064009

\bibitem{1985:CPR:Monaghan}
Monaghan J 1985 {\em Comput. Phys. Rep.\/} {\bf 3} 71--124

\bibitem{2002:JCP:Parshikov}
Parshikov A~N and Medin S~A 2002 {\em J. Comp. Phys.\/} {\bf 180} 353--382

\bibitem{2017:LJM:Dyachkov}
Dyachkov S~A, Egorova M~S, Murzov S~A, Parshikov A~N and Zhakhovsky V~V 2017
  {\em Lobachevskii J. Math.\/} {\bf 38} 893--897

\bibitem{kanel:2015}
Kanel G~I, Savinykh A~S, Garkushin G~V and Razorenov S~V 2015 {\em JETP
  letters\/} {\bf 102} 548--551

\bibitem{ashitkov:2016}
Ashitkov S~I, Komarov P~S, Ovchinnikov A~V, Struleva E~V and Agranat M~B 2016
  {\em JETP Letters\/} {\bf 103} 544--548

\end{thebibliography}

\providecommand{\newblock}{}

\end{document}